\documentclass[11pt]{article}
\usepackage{amsmath,latexsym,amssymb,epsfig,psfrag,subfigure,graphicx,verbatim,multirow,appendix}
\makeatletter
\long\def\@caption#1[#2]#3{\par\addcontentsline{\csname
  ext@#1\endcsname}{#1}{\protect\numberline{\csname
  the#1\endcsname}{\ignorespaces #2}}\begingroup
    \small
    \@parboxrestore
    \@makecaption{\csname fnum@#1\endcsname}{\ignorespaces #3}\par
  \endgroup}
\makeatother
\newcommand{\newc}{\newcommand}
\newc{\lat}{{\ell at}}
\newc{\one}{{\bf 1}}
\newc{\mgut}{M_{\rm GUT}}
\newc{\mzero}{m_0}
\newc{\mhalf}{M_{1/2}}
\newc{\five}{{\bf 5}}
\newc{\fivebar}{{\bf\bar 5}}
\newc{\ten}{{\bf 10}}
\newc{\tenbar}{{\bf\bar{10}}}
\newc{\sixteen}{{\bf 16}}
\newc{\sixteenbar}{{\bf\bar{16}}}
\newc{\gsim}{\lower.7ex\hbox{$\;\stackrel{\textstyle>}{\sim}\;$}}
\newc{\lsim}{\lower.7ex\hbox{$\;\stackrel{\textstyle<}{\sim}\;$}}
\newc{\gev}{\,{\rm GeV}}
\newc{\mev}{\,{\rm MeV}}
\newc{\ev}{\,{\rm eV}}
\newc{\kev}{\,{\rm keV}}
\newc{\tev}{\,{\rm TeV}}
\newc{\mz}{m_Z}
\newc{\mw}{m_W}
\newc{\mpl}{M_{Pl}}
\newc{\mh}{m_h}
\newc{\mA}{m_A}
\newc{\tr}{\mbox{Tr}}

\newc\CO{\order}
\newc\CL{{\cal L}}
\newc\CY{{\cal Y}}
\newc\CH{{\cal H}}
\newc\CM{{\cal M}}
\newc\CF{{\cal F}}
\newc\CD{{\cal D}}
\newc\CN{{\cal N}}
\newc{\eps}{\epsilon}
\newc{\re}{\mbox{Re}\,}
\newc{\im}{\mbox{Im}\,}
\newc{\invpb}{\,\mbox{pb}^{-1}}
\newc{\invfb}{\,\mbox{fb}^{-1}}
\newc{\yddiag}{{\bf D}}
\newc{\yddiagd}{{\bf D^\dagger}}
\newc{\yudiag}{{\bf U}}
\newc{\yudiagd}{{\bf U^\dagger}}
\newc{\yd}{{\bf Y_D}}
\newc{\ydd}{{\bf Y_D^\dagger}}
\newc{\yu}{{\bf Y_U}}
\newc{\yud}{{\bf Y_U^\dagger}}
\newc{\ckm}{{\bf V}}
\newc{\ckmd}{{\bf V^\dagger}}
\newc{\ckmz}{{\bf V^0}}
\newc{\ckmzd}{{\bf V^{0\dagger}}}
\newc{\X}{{\bf X}}
\newc{\bbbar}{B^0-\bar B^0}

\newc{\sgn}{\mbox{sgn}\,}
\newc{\m}{{\bf m}}
\newc{\msusy}{M_{\rm SUSY}}
\newc{\munif}{M_{\rm unif}}
\newc{\slepton}{{\tilde\ell}}
\newc{\Slepton}{{\tilde L}}
\newc{\sneutrino}{{\tilde\nu}}
\newc{\selectron}{{\tilde e}}
\newc{\stau}{{\tilde\tau}}
\newc{\vckm}{V_{\!\mbox{\tiny CKM}}}
\def\bar#1{\overline{#1}}
\def\vev#1{\left\langle #1 \right\rangle}

\def\beq{\begin{equation}}
\def\eeq{\end{equation}}
\def\bea{\begin{eqnarray}}
\def\eea{\end{eqnarray}}
\newc{\ie}{{\it i.e.}}          \newc{\etal}{{\it et al.}}
\newc{\eg}{{\it e.g.}}          \newc{\etc}{{\it etc.}}
\newc{\cf}{{\it c.f.}}
\def\Dsl{\,\raise.15ex\hbox{/}\mkern-13.5mu D} 
\def\delsl{\raise.15ex\hbox{/}\kern-.57em\partial}
\def\Ksl{\hbox{/\kern-.6000em\rm K}}
\def\Asl{\hbox{/\kern-.6500em \rm A}}
\def\Dsl{\hbox{/\kern-.6000em\rm D}} 
\def\Qsl{\hbox{/\kern-.6000em\rm Q}}
\def\gradsl{\hbox{/\kern-.6500em$\nabla$}}
\begin{document}
\begin{titlepage}
\begin{center} ~~\\
\vspace{0.5cm} 
\Large {\bf\Large A Gauge-Mediated Embedding of the S-MSSM} 
\vspace*{1.5cm}

\normalsize{
{\bf Antonio Delgado\footnote[1]{antonio.delgado@nd.edu},
 Christopher Kolda\footnote[2]{ckolda@nd.edu}, 
 J.~Pocahontas Olson\footnote[3]{jspeare@nd.edu}, \\ and 
 Alejandro de la Puente\footnote[4]{adelapue@nd.edu}} } \\

\smallskip  \medskip
{\it Department of Physics, University of Notre Dame,}\\
{\it Notre Dame, IN 46556, USA}

\medskip

\vskip0.6in 

\end{center}

\centerline{\large\bf Abstract}
\vspace{.5cm}
\noindent
We embed the S-MSSM  -- a singlet extension of the minimal supersymmetric standard model with an explicit $\mu$-term and supersymmetric mass term -- in a gauge-mediated supersymmetry breaking scheme.  We find that by absolving the singlet of its responsibility for solving the $\mu$-problem, we are instead able to solve the little hierarchy problem. Specifically, we find that even with a minimal embedding of the S-MSSM into a gauge-mediated scheme, we can easily raise the lightest Higgs mass above 114 GeV, while keeping top squarks below the TeV scale, maintaining perturbative unification of the gauge couplings, and without tuning the other parameters of the model. 

\vspace*{2mm}
\end{titlepage}

\section{\Large Introduction}
\label{sec:Intro}
While there are several arguments that may lead one to expect supersymmetry (SUSY) as a weak-scale extension of the Standard Model, none is more powerful than the cancellation of the quadratic divergences that generate the so-called ``hierarchy problem" of the Standard Model. SUSY avoids the order-by-order fine-tuning that one imposes on the Higgs mass of the Standard Model in favor of a doubling of the particle spectrum. This doubling is not without cost, as several problems that are resolved naturally within the Standard Model ({\it e.g.}, conservation of baryon and lepton number, suppression of flavor-changing neutral currents) require additional symmetries or structure in order to be accommodated within the minimal SUSY Standard Model (MSSM). 

In order for SUSY to cleanly solve the hierarchy problem, the new SUSY partners must lie close to the weak scale -- as the mass scale of the superpartners grows, fine-tunings of $O(m_W^2/M_{\rm SUSY}^2)$ reappear. At present, this problem is somewhat academic since none of the superpartner masses have been measured. But constraints from experiment have begun to force the MSSM into regimes in which some tuning is required. The worst of these tunings comes from the LEP lower bound on the Standard Model Higgs mass of $114\gev$~\cite{LEPHiggs}. This constraint is a problem for the MSSM for two reasons. First, in large regions of parameter space, the lightest scalar Higgs boson, $h^0$, behaves indistinguishably from the Standard Model Higgs, and so for these Higgs bosons the LEP bound applies. Second, within the MSSM, the mass of $h^0$ is bounded from above at tree level by the $Z^0$ mass, $\mz$, and so must be lifted above the LEP bound by large radiative corrections.

The MSSM does allow for large radiative corrections to the lightest Higgs mass, but the corrections necessary in order to evade the LEP bound require that the top squarks become quite heavy and/or that the trilinear stop mixing term ($A_t$) take particular values which are themselves very large. These large stop masses/stop mixing terms feed back into the mass parameters of the Higgs potential, pulling the Higgs vacuum expectation value (vev) up to scales above a TeV. Stabilizing the weak scale at $v=174\gev$ requires tunings of better than 1\% amongst the parameters of the Higgs potential. This smaller version of the original hierarchy problem is aptly named the ``little hierarchy problem" and is quite generic within the MSSM~\cite{little}.

In the last decade, 
a great deal of work has been done on finding extensions to the MSSM which naturally evade the LEP bound without requiring new tunings. These extensions include: models in which the SUSY Higgs evades experimental discovery and thus can lie below the LEP bound~\cite{gunion}; models in which the cut-off scale is explicitly made to lie near the weak scale~\cite{lowcutoff} or in which the cut-off is lowered implicitly by imposing strong coupling on the theory~\cite{fathiggs}; models in which new operators~\cite{newops,dst} or symmetries~\cite{newsyms} are grafted onto the MSSM; and models in which new particles are added to the spectrum of the MSSM. This last category divides neatly into two very different ideas. In one class are those models which add matter fields with the goal that these fields couple to the Higgs sector and generate new radiative corrections to the Higgs mass~\cite{newsyms,newmatter}. The second, and most popular, idea is to add fields to the Higgs sector itself, in order to generate new quartic terms in the Higgs potential, pushing up the Higgs mass already at tree level~\cite{quiesp}. 

The simplest implementation of this last approach is the well-studied Next-to-Minimal SUSY Standard Model (NMSSM). The NMSSM extends the spectrum of the MSSM by a singlet superfield, $S$, and then imposes an additional $Z_3$ symmetry on its superpotential. This symmetry, under which all superfields are singly charged, forbids the usual $\mu$-term of the MSSM, but replaces it with an interaction among the $S$ and Higgs fields: 
\begin{equation}
W_{\rm NMSSM} = W_{\rm Yukawa} + \lambda S H_u H_d + \frac{\kappa}{3} S^3,
\label{eq1}
\end{equation}
where $W_{\rm Yukawa}$ represents the couplings of $H_u$ and $H_d$ to the quark and lepton superfields, and $\lambda$ and $\kappa$ are dimensionless coupling constants. (Here we take the SU(2) contraction $H_u H_d\equiv \epsilon_{ij} H_u^i H_d^j=H_u^+ H_d^- - H_u^0 H_d^0$.)
When the scalar component of $S$ receives a vev, a $\mu$-term is dynamically generated.  Because the singlet vev is generated by soft SUSY-breaking mass terms in the scalar potential, it is naturally of the order of the soft masses and therefore of the electroweak scale, solving the ``$\mu$ problem" of the MSSM. (See Ref.~\cite{nmssmreview} for an extensive review of the NMSSM.) However the NMSSM has a number of difficulties and its phenomenologically viable parameter space is highly constrained both by experiment and theory, some of which we discussed in a previous paper~\cite{smssm}. In particular, it is difficult for the single $S$-field to solve both the $\mu$-problem and the little hierarchy problem simultaneously, though it can be done with careful choices of parameters.

In our previous paper~\cite{smssm}, we presented an extension of the NMSSM in which we absolved the singlet of its responsibility for solving the $\mu$-problem and found that it then solved the little hierarchy problem in a very natural way. We called this model the S-MSSM, since it is the most general version of the MSSM with one singlet superfield added\footnote{We did impose $R$-parity on the model, so more correctly, it is the most general singlet extension of the $R$-parity-conserving MSSM.}. The most important element in the S-MSSM was the inclusion of explicit masses for the $H_uH_d$ bilinear and the $S$ superfield in the superpotential:
\begin{equation}
W_{\rm SMSSM} \supset \mu H_u H_d + \frac12 \mu_s S^2.
\end{equation}

In contrast to the MSSM, the S-MSSM was shown to raise the mass of the light Higgs boson above the LEP bound at tree-level without the need of a heavy stop spectrum. After including the full one-loop effective potential and the leading two-loop corrections, we showed that the S-MSSM handily produces light Higgs masses above the LEP bound and even as high as $140\gev$. In order to generate these large corrections, $\mu_s$ must be a few times heavier than the other mass scales in the theory: for smaller $\mu_s$, the singlet mixes too much into the light Higgs, pulling down its mass; for large $\mu_s$, the singlet decouples, leaving behind the MSSM and its too-low Higgs masses. We also showed that Higgs masses above the LEP bound could be obtained ``naturally" with stop masses all the way down to $400\gev$.

But one must be careful discussing naturalness within a completely low-energy effective theory, such as the S-MSSM. Parameter choices that appear natural at low energies may turn out to be quite unnatural once that low-energy model is embedded in an ultraviolet theory. Conversely, choices that appear unnatural at low energies may simply be clues about the structure of a perfectly natural ultraviolet completion. In either case, it is difficult to speak of naturalness without a more complete theory. In Ref.~\cite{smssm} we used the stop masses as a measure of fine-tuning, arguing that if stop masses were significantly smaller than $1\tev$ and there were no other obvious tunings required to make the model work, then the S-MSSM is in some sense natural. In this paper, we will explore whether the parameter choices made in Ref.~\cite{smssm} could be generated within a more complete model, one in which the soft SUSY-breaking masses are determined by only a few inputs.

In order to do so, we will take a very natural model of SUSY-breaking, namely gauge-mediated SUSY breaking (GMSB), and use the relations of GMSB to constrain our parameter space. We will show that one still finds large, and apparently natural, regions of parameter space in which the lightest Higgs lies above the LEP mass bound. In the process we will see that the S-MSSM, by not trying to do as much as the NMSSM, can be made to fit into a GMSB scheme in more minimal and natural ways, with a much larger parameter space available in which to find phenomenologically viable models.

Our paper is organized as follows: In Section 2, we will revisit the S-MSSM discussed in Ref.~\cite{smssm}. In Section 3 we will review GMSB and the various ways in which it is implemented in the NMSSM. We then explain how the S-MSSM is can be embedded in GMSB. In Section 4 we show the results of a study of the parameter space of the GMSB-embedded S-MSSM, with a discussion in the final section. At the end of the paper is an Appendix in which the details of the S-MSSM, its potential and Higgs spectrum are laid out for the reader.

\section{\Large The S-MSSM as a Low-Energy Model}
\label{sec:LowEnergyModel}

The basic ingredients of the S-MSSM were laid out in Ref.~\cite{smssm}; we will review them here, with additional details in the Appendix.

The most general superpotential one can write for the MSSM with one additional singlet, and which preserves $R$-parity, is:
\beq \label{Wfull}
W=W_{\rm Yukawa} + (\mu+\lambda S) H_u H_d + \frac{1}{2}\mu_s S^2 + \frac{1}{3}\kappa S^3 + \xi S.
\eeq
This superpotential contains explicit mass terms both for the $H_uH_d$ pair (the usual $\mu$-term) and for the singlet itself ($\mu_s$), as well as a tadpole term for the singlet and a cubic self-coupling. 

A few comments are now in order. First, in our analysis of the S-MSSM, we will allow $\mu$ and $\mu_S$ to take arbitrary values at or below the TeV scale, independent of the vev of $S$, and so we give up any attempt at solving the $\mu$-problem (or $\mu_s$-problem). This is the key trade-off that this model makes: give up the solution to the $\mu$-problem in the hope of solving the little hierarchy problem. Of course, one could imagine ways of solving the $\mu$-problem with additional fields or symmetries, but we are only interested in the effective theory described by our superpotential above.

Second, a bare tadpole term can be removed by field redefinitions, and even though there is no symmetry that explicitly forbids it, the non-renormalization theorem will prevent the tadpole term from being generated radiatively until SUSY is broken. Once SUSY is broken, the tadpole could reappear, but not in a calculable way. If the tadpole coefficient $\xi\gg M_{\rm SUSY}$, then the singlet will develop a huge vev and ruin the electroweak symmetry breaking; if the S-MSSM is to correctly model physics at the electroweak scale, this cannot be the case and so we do not consider it. If $\xi\ll M_{\rm SUSY}$, then the tadpole is irrelevant to our discussion, and we again ignore it. Only if $\xi\sim M_{\rm SUSY}$ can the tadpole be relevant, but since this requires some tuning of the model, and since we don't need the tadpole to play any particular role in the model, we do not consider it further. 

Third, while the $S^3$ term is required in the usual NMSSM in order to stabilize the potential in the $S$ direction, our potential is stabilized by the explicit mass term, $\mu_s$. Because the $S^3$ term is no longer required, and because its effects tend to be small anyway, we will take $\kappa$ to be effectively zero in the analysis that follows. One could reanalyze this model for non-zero $\kappa$, but we don't expect that there is much to be gained in this direction.

Therefore our superpotential is reduced to:
\begin{equation}\label{W}
W_{\rm SMSSM}=W_{\rm Yukawa} + (\mu+\lambda {S}) H_{u}{H_{d}}+\frac{1}{2}\mu_{s}{S}^{2}.
\end{equation}
We call the model represented by this superpotential the S-MSSM.
Philosophically, the S-MSSM is an attempt to {\it barely}\/ extend the MSSM; not only is the particle content minimally extended, but the structure of the vacuum will be nearly identical to that of the MSSM. In this sense, our philosophy is similar to that in Refs.~\cite{dst,chacko}. The authors of Ref.~\cite{dst} do briefly considered a model like the S-MSSM, though without examining its implications in any detail. The authors of Ref.~\cite{chacko} examine a class of models which overlaps the one being considered here.

The superpotential in Eq.~(\ref{W}), along with the soft-breaking terms, yields the following potential for the neutral Higgs and singlet fields:
\begin{eqnarray}
\label{eq:Potential}
V_H^{(0)}  &=&  ( m_{H_u}^2 + |\mu + \lambda S|^2 ) | H_u^0 |^2 +
( m_{H_d}^2 + |\mu + \lambda S|^2 ) | H_d^0 |^2 + m_s^2 |S|^2 \nonumber \\
  & & {}+\,\, | \mu_s S - \lambda H_u^0 H_d^0 |^2
+ \left( B_s S^2 - ( B_\mu + \lambda A_{\lambda} S ) ( H_u^0 H_d^0 ) + h.c. \right) \nonumber \\
&& {}+\,\, \frac{1}{8} ( g^2 + g'^2 ) ( |H_d^0|^2 - |H_u^0|^2 )^2 
\end{eqnarray}
where $g$ and $g^\prime$ are the gauge couplings of $SU(2)_W$ and $U(1)_Y$.  

Minimizing the potential with respect to $H_u$, $H_d$ and $S$ yields three constraints (details can be found in the Appendix):
\begin{eqnarray}
\label{eq:m_Z}     
\frac{1}{2} m_Z^2 &=& \frac{ m^2_{H_d} - m^2_{H_u} \tan^2\beta }{ \tan^2\beta-1 } - \mu_\text{eff}^2 \\
\label{eq:sin2Beta}
\sin 2\beta &=& \frac{2 B_{\mu,\text{eff}}}{m^2_{H_u} + m^2_{H_d} + 2\mu_\text{eff}^2 + \lambda^2 v^2} \\
\label{eq:Svev}
v_s &=& \frac{\lambda v^2}{2}\, \frac{(\mu_{s}+A_{\lambda})
\sin 2\beta-2\mu}{ \mu_{s}^{2}+\lambda^{2}v^{2}+m_{s}^{2}+2B_{s}}\quad \xrightarrow[\mu_s\to\infty]{}\quad \frac{\lambda v^2}{2\mu_s}\sin 2\beta, \label{vs}
\end{eqnarray}
where $v_s= \left \langle S \right \rangle$ and $v_{u,d}= \left \langle {H_{u,d}} \right \rangle$, with $v=(v_u^2+v_d^2)^{1/2}=174\gev$ and
\begin{eqnarray}
\mu_\text{eff} &\equiv& \mu + \lambda v_s, \\
B_{\mu,\text{eff}} &\equiv& B_\mu + \lambda v_s (\mu_s+A_\lambda ).
\end{eqnarray}

There are several interesting limits in this model.  The first is the limit in which $\mu_s$ is large compared to the other masses in the model. In this case, the potential becomes quite steep along the singlet direction, forcing the vev of $S$ to lie near the origin while  decoupling the singlet from the theory, leaving only the MSSM as the effective theory.
We could also consider the case in which $\mu_s$ is small.  In this regime, the singlet mixes strongly with the CP-even Higgs bosons, decreasing the lightest Higgs boson mass, and thus making it harder to surpass the LEP mass bound. However, one may be able to find regions of parameter space in which the light Higgs states are effectively hidden from detection at LEP, thus evading the bound in a way similar in spirit to the proposal of Dermisek and Gunion~\cite{gunion}, though we will not discuss that possibility here.

The S-MSSM works to maximize the light Higgs mass precisely between the two limits discussed above. If $\mu_s$ is too large, the MSSM predictions are recovered; if $\mu_s$ is too small, the mixing drives down the Higgs mass. This can all be seen most easily by expanding the Higgs masses of the S-MSSM in powers of $1/\mu_s$ (see the Appendix for full details):
\begin{eqnarray}
\label{eq:masses1}
m_{A^0_1}^2 & \simeq & \frac{2 B_\mu}{\sin2\beta} + \frac{4A_{\lambda}\lambda^{2}v^{2}}{\mu_s} - \frac{2\mu\lambda^2 v^2}{\mu_s\sin2\beta}  \\
\label{eq:masses2}
m_{A_2^0 , H_2^0}^2 & \simeq & \mu_s^2 + (2 \lambda^2 v^2 + m_s^2 \mp 2 B_s )   \\
\label{eq:masses3}
m_{h^0 , H_1^0}^2 & \simeq & 
	\left.m^2_{h^0,H^0_1}\right|_{\mbox{\scriptsize MSSM}}
	+ \frac{2 \lambda^2 v^2}{\mu_s} \left(\mu \sin2\beta - A_\lambda \mp \Delta\right)  
\end{eqnarray}
where the first term in Eq.~(\ref{eq:masses3}) represents the masses of the lightest CP-even Higgs bosons, as calculated within the MSSM:
\begin{eqnarray}
\label{eq:MSSMmh}
\left.m^{2}_{h^0,H^0_1}\right|_{\mbox{\scriptsize MSSM}} &\equiv& 
	\frac{1}{2} \left( m_{A_1^0}^2 + m_Z^2 \mp
	\sqrt{ ( m_{A_1^0}^2 + m_Z^2 )^2 - 4 m_{A_1^0}^2 m_Z^2 \cos^2 2 \beta }  \right),
\end{eqnarray}
and $\Delta$ corresponds to the $O(1/\mu_s)$ correction to the splitting of the scalar Higgs masses:
\begin{equation}
\label{eq:deltaCorrection}
\Delta \equiv 
\frac{A_{\lambda}(m^{2}_{Z}-m^{2}_{A_1^0})\cos^{2}2\beta-\mu(m^{2}_{A_1^0}+m^{2}_{Z})\sin2\beta}{\sqrt{(m^{2}_{A_1^0}+m^{2}_{Z})^{2}-4m^{2}_{A_1^0}m^{2}_{Z}\cos^{2}2\beta}}.
\end{equation}
It is obvious from Eqs.~(\ref{eq:masses1})-(\ref{eq:masses3}) that at very large $\mu_s$, $S\simeq H^0_2+i A^0_2$, both of which have masses $\sim\mu_s$, and the remaining Higgs bosons behave as in the MSSM.

It is also helpful to consider the lightest Higgs mass in the Higgs decoupling limit, in which $m_{A_1^0}$ is taken large, since this simplifies our expressions considerably, while also maximizing the light Higgs mass:
\begin{eqnarray}
\label{eq:largemA}
	m_{h^0}^2 &\simeq &
               m_Z^2 \cos^2 2 \beta + \frac{2 \lambda ^2 v^2}{\mu_s }  	
			\left( 2 \mu \sin 2 \beta 
				- A_\lambda \sin^2 2 \beta \right) 
		- \frac{\lambda^2 v^2}{\mu_s^2}
	\Big\lbrace 4\mu (\mu -A_\lambda \sin 2\beta) \nonumber\\
&& \,\,\,{}+	 (A_\lambda^2 - 3\lambda^2 v^2 - m_s^2 - 2 B_s)\sin^2 2\beta	
	  + m_Z^2 \sin^2 2\beta \cos^2 2\beta \Big\rbrace.  
\end{eqnarray}
Here we have included both the leading and sub-leading terms in the $1/\mu_s$ expansion, both of which are required to reproduce many of our results. (Our numerical work in the next section is done exactly, however.) 

In principle, one can keep increasing $\lambda$ to obtain arbitrarily large Higgs masses. However, if we take the apparent gauge coupling unification in the MSSM as a sign that physics must remain perturbative up to scales $\sim 10^{16}\gev$, then at the weak scale we must have $\lambda \lsim 0.7$; see the Appendix for more details. Choosing values for $\lambda>0.7$ is equivalent to placing a new and lower cut-off on the theory, an avenue for raising the Higgs mass that is already known to work.

Using this model with $\lambda$ bounded by perturbativity, and a simplified set of input parameters, we showed in Ref.~\cite{smssm} that the S-MSSM can generate very large Higgs masses (up to $140\gev$) without any apparent tunings, apart from those already implicit in the $\mu$-problem. The model itself is surprisingly similar to the MSSM, and because we find $\mu_s\gsim 1\tev$ in our regions of interest, the low-energy spectrum of the S-MSSM will be that of the MSSM. The only readily observable difference will be that the light Higgs mass in the S-MSSM will be substantially larger than one would expect given the masses of the stops, which will presumably be measured at the LHC.

In the next section, we will embed the S-MSSM into a model of gauge-mediated SUSY breaking, in order to explore its parameter space in a more motivated and complete way.


\section{\Large Gauge Mediation in Models with a Singlet}
Gauge-mediated supersymmetry breaking is a robust scheme for generating the soft masses within low-energy models of SUSY, with a very small number of free parameters determining the entire spectrum for the model. (See Ref.~\cite{GaugeMedReview} for a review.) In GMSB, the effects of SUSY breaking are mediated through messenger fields $\{\Phi,\bar\Phi\}$ interacting with the MSSM through gauge interactions, thus providing a natural solution for the SUSY flavor problem. The messenger fields transform as a $\bf{5}$ and $\bf{\bar{5}}$ of $SU(5)$, thus preserving the appearance of gauge coupling unification that partially motivates our interest in SUSY. In GMSB theories, one also implements a hidden sector whose main effects are parametrized by a chiral superfield/spurion $X$ which acquires a non-vanishing vev in both its scalar and $F$-components: $\vev{X} = M+\theta^2 F$ (we will assume $F\ll M^2$). In its minimal form, GMSB then couples the $X$ superfield to the messenger fields in the superpotential, but not to any of the usual MSSM fields:
\begin{equation}
\label{minForm}
	W=X\Phi\bar{\Phi}.
\end{equation}
The vev of $X$ yields mass terms for the messenger fields: the scalar messengers acquire masses $M\pm\sqrt{F}$ while the fermionic messengers have mass $M$, breaking SUSY in the messenger sector. SUSY breaking is then mediated to the observable sector by gauge interactions at one- and two-loops. Specifically, soft masses for gauginos and sfermions are generated at one- and two-loops respectively. At the messenger scale, $M$, these are given by
\begin{equation}
\label{SBgauginos}
M_{i}(M)=n\frac{\alpha_{i}}{4\pi}\frac{F}{M}
\end{equation}
for gauginos, and 
\begin{equation}
\label{SBsfermions}
m_{\tilde{f}}^{2}(M)=2n\sum_{i}C^f_i\frac{\alpha_{i}^{2}}{16\pi^{2}}\Big(\frac{F}{M}\Big)^{2}
\end{equation}
for scalars, where $n$ specifies the number of messenger pairs, $i$ indexes the gauge group, and $\alpha_{i}$ and $C^f_i$ are the gauge coupling and quadratic Casimir associated with each group, $i$, and chiral superfield, $f$.
In GMSB, contributions to the trilinear $A$-terms appear first at two loops, while comparable mass terms appear at one loop, so these can be simply set to zero at the messenger scale.

So far, no mention has been made of the $\mu$ and $B_\mu$ parameters in GMSB. There are typically two approaches one can take here. The first is to derive $\mu$ and $B_\mu$ from the conditions of electroweak symmetry breaking, putting them into the superpotential and soft-breaking Lagrangian by hand. In doing so, one implicitly assumes that the physics responsible for generating $\mu$ and $B_\mu$ can be segregated from the physics which generates the remainder of the soft-breaking terms. Alternatively, one can extend the model to produce the $\mu$ and/or $B_\mu$ dynamically, but the numerous difficulties that arise in this approach have come to be called the $\mu/B_\mu$-problem of GMSB. For this reason, phenomenological studies of the gauge-mediated MSSM usually favor the first approach.

The problem is quite different in the NMSSM. Here one has no choice but to solve the $\mu$- and $B_\mu$-problems since they arise dynamically from the minimization of the scalar potential: $\mu=\lambda v_s$ and $B_\mu=\lambda A_\lambda v_s$. The literature on embedding the NMSSM into GMSB is quite involved, but we now review a few of the leading ideas in order to give the reader a flavor of the issues and tunings that arise.

The minimal GMSB version of the NMSSM is nothing more than the model described above in Eqs.~(\ref{minForm}) -- (\ref{SBsfermions}): a single spurion which communicates SUSY-breaking to a set of messengers charged under $SU(3)\times SU(2)\times U(1)$, who then communicate it to the gauginos and scalars at one and two loops respectively. The superpotential is that of the NMSSM plus the minimal GMSB superpotential of Eq.~(\ref{minForm}):
\begin{equation}
W=\lambda SH_{u}H_{d}+\frac{1}{3}\kappa S^{3}+X\Phi\bar{\Phi}.
\end{equation}
Unfortunately, the $S$-scalar, being neutral under the SM gauge groups, does not receive a soft-breaking mass, $m_s^2$, until 3-loops. Because $m_s^2$ sets the scale for the $S$ potential, $v_s$ is suppressed, too small to generate the $\mu$-term required by the chargino and neutralino experimental mass bounds~\cite{Murayama}. A closely related problem is that the superpotential above possesses an approximate Peccei-Quinn/$R$-symmetry under which all the MSSM superfields, as well as $S$ carry charge +2/3. This symmetry is explicitly broken by $A$-terms, and spontaneously broken when the singlet gets a vev. But because the $A$-terms are small within gauge mediation, the pseudoscalar Higgs behaves as an axion and is too light to be phenomenologically viable. Finally, this model possesses a $Z_3$ symmetry which is broken spontaneously at the weak scale, seemingly generating domain walls in the early universe that would be difficult to inflate away. 
 
Because the minimal embedding of the NMSSM into gauge mediation is so fraught with difficulties, several non-minimal embeddings have been proposed which are phenomenologically viable in at least some portion of their parameter space, and which manage to get light Higgs masses above the LEP bound, though at some cost. We will quickly review two such proposals.

The first proposal couples $S$ directly to the messenger fields in the superpotential: 
\begin{equation}
W=\lambda SH_{u}H_{d}+\frac{1}{3}\kappa S^{3}+X\Phi\bar{\Phi}-\eta S\Phi\bar{\Phi}.
\label{idea2}
\end{equation}
Here a soft mass-squared for the singlet is generated already at one loop, but requires extra insertions of the SUSY-breaking mass scale, producing a mass-squared for $S$,
\begin{equation}
m_{s}^{2}=\frac{\eta^{2}}{4\pi^{2}}\frac{F^3}{M^4},
\end{equation}
which is too small to produce the required $\mu$-term. Thus the model would seem to fail just as the minimal case. However, as pointed out in Ref.~\cite{Ellwanger_2}, $S$ must share the same symmetry properties as the spurion, $X$, in Eq.~(\ref{idea2}). Thus there can be tadpoles in the model which are cut off by the messenger scale and which could drive $S$ to get a vev. There are two possible tadpoles, one in superpotential, $W$, and one in the potential, $V$, which scale as:
\bea
\xi_W &\sim& \frac{\eta}{8\pi^2} F\log(\Lambda^2/M^2) \\
\xi_V &\sim& \frac{\eta}{16\pi^2} \frac{F^2}{M}
\eea
where $\Lambda$ is some cut off, presumably near $M$. Of course, the precise values of the tadpole contributions depend on physics at very high scales, but if we take these expressions literally, one can choose $F$, $M$ and $\eta\ll1$ so that both tadpoles are of order the weak scale, driving $S$ to get a vev of the same size. This model produces Higgs masses above the LEP bound either at low $\tan\beta$, large $\lambda$ and very low messenger scales, or at large $\tan\beta$ with very large $M$ and very heavy stops ($\gsim 2\tev$). In the first case, perturbative unification of the gauge couplings at $10^{16}\gev$ is not preserved without some additional physics at intermediate scales~\cite{Ellwanger_2}.
 
One variation on the previous model is to allow for two sets of messenger fields:
 \begin{equation}
 W=\lambda SH_{u}H_{d}+\frac{1}{3}\kappa S^{3}+X(\Phi_{1}\bar{\Phi}_{1}+\Phi_{2}\bar{\Phi}_{2})-\chi S\Phi_{2}\bar{\Phi}_{1}.
 \end{equation}
Notice that $S$ couples only to the bilinear $\Phi_2\bar\Phi_1$, a requirement enforced by a $Z_3$ symmetry; this same symmetry also forbids a tadpole in $S$. The non-diagonal couplings to the messengers prevents $m_s^2$ from arising until two loops, producing a soft-breaking mass, and thus a $\mu$-term, of the right order~\cite{DelgadoGaugeMed}. As in the usual GMSB scenarios, $A$-terms for the squarks are not generated until three loops, and are therefore quite small.

This model was studied in detail in Ref.~\cite{DelgadoGaugeMed}, where the authors found two limiting behaviors for the light Higgs. In one case the light Higgs was essentially MSSM-like, with a tree-level mass below the $Z^0$. Here the usual MSSM radiative corrections can lift the Higgs mass, but because this is GMSB, the $A_t$-term is nowhere close to the maximal mixing case, and so one must resort to stop masses around $2\tev$ in order to beat the LEP bound. In the second limit, corresponding to large $\lambda$, $A_\kappa$, and $|m_s^2|$, the light Higgs mass can be raised above the LEP bound by the presence of the singlet, but not by significant amounts. This second case lives right on the edge of the allowed parameter space, with couplings that are almost non-perturbative.  

For the S-MSSM, we will accept the philosophy adhered to by most implementations of gauge mediation in the MSSM, namely, we will assume the minimal gauge mediation sector, with no direct coupling of the singlet to the messengers:
\begin{equation}
W=(\mu+\lambda S)H_{u}H_{d}+\frac{1}{2}\mu_s S^{2}+X\Phi\bar{\Phi}.
\end{equation}
As noted above, for the NMSSM this does not work, because it generates too small of a soft mass for the singlet and therefore too small of a $\mu$-term; in our case, a small $m_s^2$ is no longer a problem, since we include an explicit $\mu$-term. We will derive the values of $\mu$ and $B_\mu$ necessary to produce electroweak symmetry breaking with the appropriate (chosen) value of $\tan\beta$. We shall take $\mu_s$ as an unknown input parameter; and because it plays almost no role whatsoever, we will set $B_s=0$. Of course, this model has a $\mu/B_\mu/\mu_s$-problem, but it is not really any worse than the MSSM when embedded in gauge mediation.

Furthermore, this potential has no symmetries, approximate, discrete or otherwise, that are broken at the weak scale. Thus there is no concern about domain walls or light axions. There {\it is}\/ a concern about generating tadpoles, but because our superpotential is technically natural until SUSY is broken, and because the resulting tadpoles are not calculable and depend on unknown details of the full, ultraviolet theory, we will assume they play no important role here.

With these caveats, we will go ahead and use GMSB to calculate the soft breaking parameters in the Lagrangian of the S-MSSM in the next section.

\section{\Large A Gauge-Mediated Embedding}

Our procedure for finding a point in the gauge-mediated S-MSSM parameter space is as follows. We begin by choosing the messenger scale, $M$. In particular, we do not want the contributions from supergravity mediation, which scale as $F/M_{\rm Pl}$ to compete with those of gauge mediation, so we insist on $M\leq 10^{13}\gev$. For each point we also choose a value of $\tan\beta$, and given that value, we set $\lambda$ equal to the largest value consistent with perturbativity all the way up to $10^{16}\gev$. (See the Appendix for details on this procedure.)

Given $\tan\beta$, we know the weak-scale values for the Yukawa couplings, and we run them, along with $\lambda$ and the gauge couplings, up to the scale $M$ using their respective renormalization group equations (RGEs); we work at one loop for soft masses, and two loops for dimensionless couplings. At $Q=M$ we choose a value of $F$ and set the soft-breaking masses for the scalars and gauginos using the boundary conditions of gauge mediation, Eqs.~(\ref{SBgauginos}) -- (\ref{SBsfermions}), with $n=1$ pair of messenger fields. We also set $A_t=A_\lambda=0$ and $m_s^2=B_s=0$ at $M$. Given these boundary conditions at $M$, we run all masses and couplings down to the weak scale. Since we are not adding any dimensionless couplings to the NMSSM, the relevant RGEs are those of the NMSSM with $\kappa=A_\kappa=0$~\cite{2loops}. The RGEs for $\mu$ and $\mu_s$ play no role in this analysis.

At the weak scale, we choose a value for $\mu$.
We then minimize the combined Higgs-singlet scalar potential using the one-loop effective potential, solving for $v$, and adjusting the value of $F$ until we obtain $v=174\gev$ or we discover that no such vacuum is allowed for that $\mu$.
Finally, from the minimization procedure we also obtain $B_\mu$ and $v_s$. We note that the low-energy values for $m_s^2$ and $B_s$ are very small and play essentially no role in the analysis.

The physical spectrum of the model is then calculated from the soft masses, including the masses and mixings among the neutralinos, charginos, and stops, all at tree level.
In the Higgs sector, the minimization is done using the complete one-loop effective potential, including the contributions from the singlet sector. The mass of the lightest scalar Higgs is then corrected using the leading two-loop contributions. We add these in quadrature to the one-loop calculation, using FeynHiggs~\cite{feynhiggs} to obtain the two-loop correction. These (negative) corrections are sizable, but since the dominant terms come from gluinos, which are largely unaffected by the addition of the singlet, the error entering our analysis from using an MSSM two-loop computation should be quite small.

For any point in the parameter space, we also apply all relevant experimental bounds. The most constraining bounds come from searches for neutralinos and charginos at LEP. Because the lightest neutralino, $\chi^0_1$, turns out to be the next-to-lightest SUSY particle (NLSP) in the regions of parameter space which we are studying, it can decay to a gravitino and photon either inside or outside the LEP detectors, depending on its lifetime. For decays outside the detector, the neutralino is simply missing energy, and the LEP bound of $46\gev$ applies. If it decays inside the detector, the LEP bound on missing energy plus hard photon demands that $m_{\chi^0_1}>96\gev$. The transition from one case to the other occurs at roughly $\sqrt{F}\sim 10^6\gev$~\cite{GaugeMedReview}, which corresponds roughly to $M\sim 10^8\gev$; for $\sqrt{F}$ below $10^6\gev$, the tighter mass constraint applies. 
The limit on the lightest chargino, $94\gev$ for $\tan \beta < 40$~\cite{delphi}, was usually redundant once the neutralino mass bound was applied.

We will show our results in several ways. We begin by considering two random scans of 10K points, one scan each with messenger scale $M=10^{10}\gev$ and $M=10^{13}\gev$, randomly varying other inputs within the ranges: 
\beq 
\begin{array}{c} 2\leq \tan\beta\leq 6 \\
2\leq \mu_s/\mu \leq 5 \\ 300\gev\leq\mu\leq 900\gev 
 \end{array} \label{scan}
 \eeq
Note that we only consider a small range of $\tan\beta$. For $\tan\beta\lsim 2$, the tree-level MSSM Higgs mass is very small and it is very hard to lift its mass above the LEP bound, even with the singlet-induced corrections. For $\tan\beta\gsim 6$, the singlet corrections disappear, since the leading contributions fall as $\sin2\beta\sim 1/\tan\beta$ for large $\tan\beta$.
We also choose $\mu>0$ since it is clear from Eq.~(\ref{eq:largemA}) that this will maximize the light Higgs mass. The upper and lower bounds are, roughly, where the LEP bound on the neutralino begin to exclude the parameter space (lower bound), and where the stops are pushed well above a TeV (upper bound).
Finally, we choose $\mu_s$ to be a multiple of $\mu$, so that $\mu_s$ is always large compared to the other mass scales present. (Choosing $\mu_s>0$ is perfectly general, since we can do so by a redefinition $S$.) We do this to ensure that the actual mixing of the singlet into the light Higgs is small, since any mixing will tend to decrease the light Higgs mass, as we showed in Ref.~\cite{smssm} and as we will see again below. The points in this parameter space are chosen randomly using a flat distribution between the limits given above.

\begin{figure}[t]
\begin{center}
\includegraphics[width=0.85\linewidth]{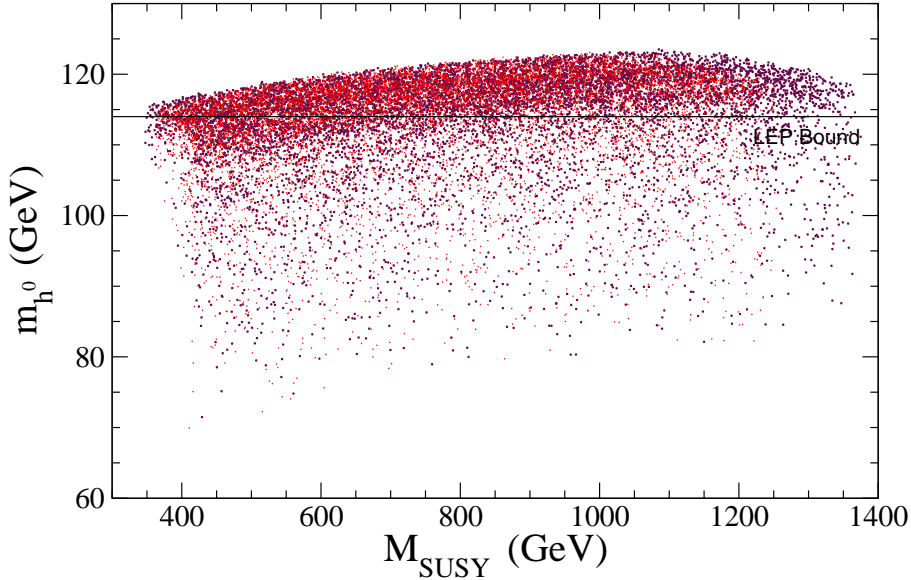}
\caption{Scatterplot of the lightest Higgs boson mass versus the effective stop mass, $M_{\rm SUSY}$, for $10^4$ randomly selected points with a messenger scale $M=10^{10}\gev$ (black) and $M=10^{13}\gev$ (red), and within the parameter space given in Eq.~(\ref{scan}). The two regions overlaps almost exactly, except at very large $M_{\rm SUSY}$. The solid black line is the LEP bound of 114 GeV.\label{fig:Scatterplot}}
\end{center}
\end{figure}

Figure~\ref{fig:Scatterplot} shows the results of this scan. Here we plot the mass of the lightest Higgs boson, calculated using the leading two-loop corrections, against an effective mass parameter, $M_{\rm SUSY}$, defined as the geometric mean of the stop masses:
\beq
M_{\rm SUSY} \equiv \left(m_{\tilde t_1} m_{\tilde t_2}\right)^{1/2}.
\eeq
The red points have $M=10^{13}\gev$, the black have $M=10^{10}\gev$.
On the plot we also show the LEP bound on the Standard Model Higgs of $114\gev$. It is immediately obvious that the great majority of the points considered lie above the LEP bound, for either messenger scale. In fact, for $M_{\rm SUSY}\simeq 400\gev$, a large percentage of the points are already consistent with all experimental limits, and by 
$M_{\rm SUSY}\simeq 500\gev$, more than half the points are consistent. Thus we see that the LEP bound can be surpassed quite naturally, without any tuning of our parameters, in the gauge-mediated S-MSSM.

It is also apparent that there is little difference between the two messenger scales. The additional running that occurs in the $M=10^{13}\gev$ case has negligible impact on the Higgs spectrum. Thus it appears that our results are fairly robust to variation of the messenger scale. This is in fact true, until the messenger scale falls below $10^8\gev$, when the experimental bound on the neutralino becomes significantly stronger due to decay inside the LEP detectors, as discussed above.

As expected, the upper bound on the Higgs mass is lower than we found in Ref.~\cite{smssm}. There we treated the S-MSSM as an effective theory, where all parameters could be chosen to maximize the Higgs mass, and we found that we could push the Higgs mass above $140\gev$. Here, with the additional structure of gauge mediation imposed on the model, we are not able to achieve Higgs masses above $124\gev$. Nonetheless, masses above $114\gev$ are quite generic and even expected within the parameter space we considered.

\begin{figure}[ptb]
\begin{center}
\includegraphics[width=0.67\linewidth,trim=10 20 70 80,clip]{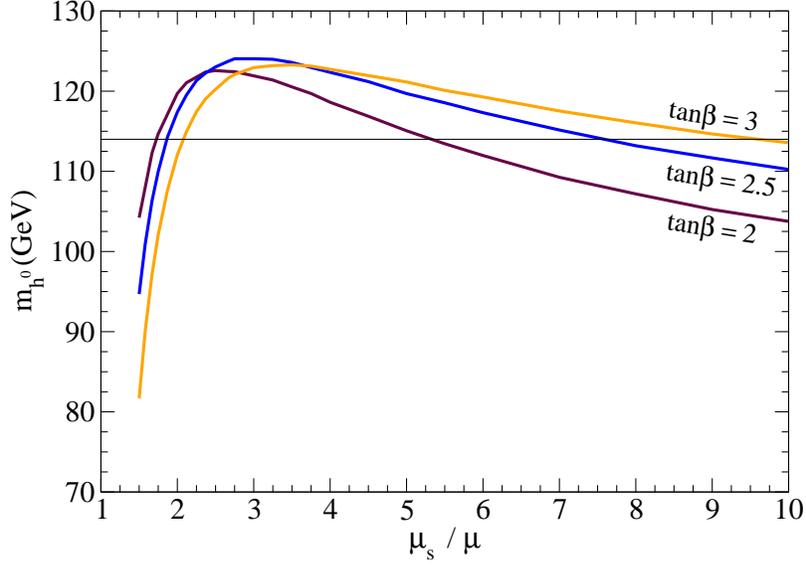}
\caption{The lightest Higgs mass as a function of $\mu_s/\mu$. For each line, the messenger scale $M=10^{13}\gev$, and $F/M$ and $\mu$ are chosen so that $m_{\tilde t_2}\simeq 1\tev$. Each line represents a different value of $\tan\beta$ as labelled. The lines end at low $\mu_s$ when electroweak symmetry breaking fails. \label{fig:PickingMuS}}
\end{center}
\end{figure}

One issue that deserves more discussion is the value of $\mu_s$. Throughout the parameter space above, we chose $\mu_s$ to be large compared to the other parameters in the Higgs potential. Clearly as $\mu_s\to\infty$, the effects of the singlet must decouple. But as discussed above, at low $\mu_s$, the singlet has a substantial mixing into the lightest Higgs, pulling down its mass. In order to obtain a large Higgs mass, we must therefore choose $\mu_s$ within some limited range, bounded from above {\it and}\/ below. This could be a potential source of tuning in our model, and so we plot the effect of changing $\mu_s$ in Figure~\ref{fig:PickingMuS}.

In this figure, we take three values of $\tan\beta=\{2,2.5,3\}$, and plot the maximum mass of the lightest Higgs as a function of $\mu_s/\mu$, requiring both stops to lie below $1\tev$. (We vary $F/M$ and $\mu$ to obtain $v=174\gev$ and $m_{\tilde t_2}=1\tev$.) It is evident from Figure~\ref{fig:PickingMuS} that at $\mu_s\lsim 2\mu$, the mixing of the singlet into the light Higgs causes a rapid decrease in the Higgs mass, pushing it below the LEP bound (solid line). However, for $\mu_s\gsim 2\mu$, there is a wide range which remains consistent with LEP. In particular, the decoupling is very slow as $\mu_s$ increases, so that even for $\mu_s\simeq 10\mu$ one can find points above the LEP bound, at least for some choices of $\tan\beta$. Thus we do not consider the choice of $\mu_s$ as introducing a new tuning into the model. Instead, the LEP bound is simply telling us that $\mu_s$ must lie above roughly $2\mu$ but must remain below a few TeV, a natural expectation within SUSY models. 

\begin{figure}[pbt]
\begin{center}
\includegraphics[width=0.67\linewidth,trim=10 20 70 80,clip]{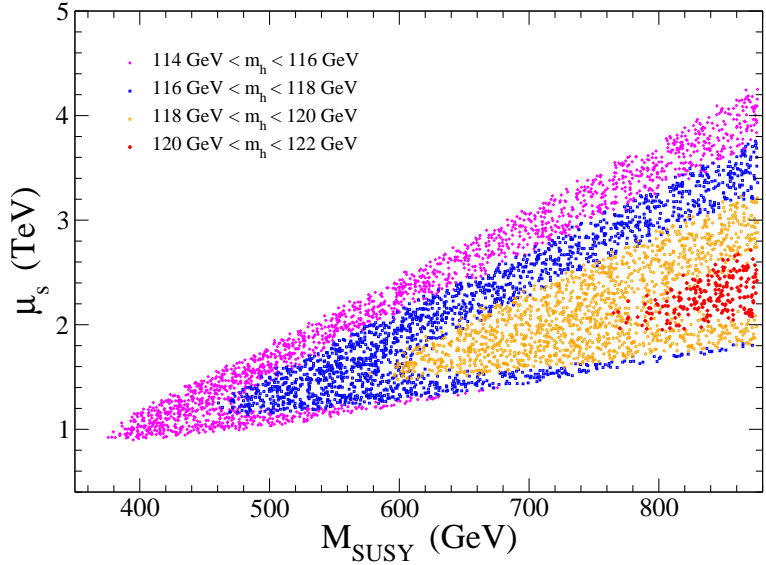}
\caption{A scan of parameter space for $\tan\beta=2$ and $M=10^{13}\gev$, varying $\mu$ and $\mu_s/\mu$ within the ranges specified in the text. All points in the figure are consistent with experimental bounds, including the bound on the Higgs mass of $114\gev$. The points are color-coded by the light Higgs mass calculated from the model parameters, in $2\gev$ steps, beginning with $m_{h^0}<116\gev$ on the outside of the triangle, and increasing to $m_{h^0}< 118$, 120, and $122\gev$ as one moves into the triangle and to the right. \label{fig:MuSPlot}}
\end{center}
\end{figure}

We can see more explicitly the correlation among $\mu_s$, $M_{\rm SUSY}$ and the Higgs mass in Figure~\ref{fig:MuSPlot}. Here we have chosen a single value of $\tan\beta=2$ and $M=10^{13}\gev$, and varied the other parameters as before. We plot the stop mass scale $M_{\rm SUSY}$ versus $\mu_s$, keeping only those points consistent with all experimental bounds, {\it including the Higgs mass bound}. We then color code the points based on the calculated Higgs mass. Notice that the Higgs mass constraint generates a triangular region in the figure. For very light stops, only a small range in $\mu_s$ is allowed, but as the stops become heavier, the range for $\mu_s$ quickly grows to allow larger values of $\mu_s$, and even heavier $h^0$. However, it does not grow to include smaller $\mu_s$ (and in fact $\mu_s\simeq 1\tev$ is no longer allowed for heavy stops though it is for light stops), an artifact of the GMSB boundary conditions, which implicitly correlate $\mu$ to the stop masses, requiring larger and larger $\mu$ as $M_{\rm SUSY}$ increases. From the color coding one also sees the behavior seen in Figure~\ref{fig:PickingMuS}, where the Higgs mass first increases with increasing $\mu_s$ but then decreases as decoupling sets in.

In order to provide the reader with more details about the spectra predicted in these models, we have chosen a number of models which were found in scans with $M=10^8$, $10^{10}$ and $10^{13}\gev$ and displayed their inputs parameters in Table~\ref{tab:SomePoints}. We also show the masses of the stops and the calculated Higgs mass. The points with the heavier stop mass just below a TeV were chosen to maximize the Higgs mass, though at the cost of heavier stops. The points with light stops were chosen to minimize the stop masses, though at the cost of lighter Higgs masses.

\begin{table}[bt]
    \begin{center}
    \begin{tabular}{|c|c|c|c||c|c||c|}
        \hline
        $M$  & $\tan \beta$ & $\mu$ & $\mu_s$ & $m_{\tilde{t}_{1,2}}$ & $m_{h^0}$  & $m_{h^0}${\small (MSSM)}  \\
        \hline \hline
	$10^8$ & 2 & 600 & 1500 & 790, 920 & 123 & 86 \\ \hline
	$10^8$ & 6 & 500 & 3000 & 760, 910 & 117 & 110   \\ \hline
	$10^{10}$ & 2 & 350 & 875 & 350, 435 & 115 & 73   \\ \hline
	$10^{10}$ & 3 & 300 & 1125 & 360, 560 & 115 & 89   \\ \hline
	$10^{10}$ & 4 & 350 & 875 & 350, 435 & 115 & 99   \\ \hline
	$10^{13}$ & 2 & 400 & 1000 & 360, 460 & 115 & 74  \\ \hline
	$10^{13}$ & 2 & 850 & 2125 & 710, 960 & 123 & 85  \\ \hline
	$10^{13}$ & 2.5 & 750 & 2060 & 730, 960 & 124 & 94  \\ \hline
	$10^{13}$ & 3 & 700 & 2450 & 730, 970& 123 & 100  \\ \hline
	$10^{13}$ & 6 & 400 & 2400 & 470, 630 & 114 & 107   \\ \hline
	$10^{13}$ & 6 & 600 & 3600 & 720, 940 & 118 & 111  \\ \hline
    \end{tabular}
    \caption{A sampling of points with Higgs masses above the LEP bound, either chosen to maximize $m_{h^0}$ or to minimize the stop masses. All masses are in GeV.     \label{tab:SomePoints}}
    \end{center}
\end{table}

For comparison, we also show the lightest Higgs mass as predicted in the MSSM using the same low-energy spectrum; specifically, we take all sparticle masses and $m_{A^0}=m_{A^0_1}$ and calculate $m_{h^0}$ in the MSSM. From the perspective of the LHC, both of these sets of models would look identical, {\it except}\/ for the greatly increased Higgs mass in the S-MSSM. In particular, notice that all the points in the table would be ruled out by the LEP bound in the MSSM alone. 

Though it is not obvious from the table, as one changes the messenger scale $M$, there is very little change in the values of the Higgs mass that one obtains for a given stop mass scale (see Figure~\ref{fig:Scatterplot}). However, this breaks down for $M$ at or below 
$10^8\gev$ case, for which we must apply the stricter mass constraint on the neutralinos, accounting for the decay of the neutralino inside the LEP detectors. For both of these reasons, models with heavy Higgs and light stops are very hard to generate for a messenger scale at or below $10^8\gev$.

For completeness, we also take one of the points from the table (the row with the lightest stops) and show its complete spectrum. Specifically, for $M=10^{13}$, $\tan\beta = 2$, $\mu=400$ GeV, $\mu_s=1$ TeV, we find $m_{h^0}=115\gev$, and: 
\beq \begin{array}{rcr}
  M_{1,2,3}=\{60,125,410\}\gev  && m_{\tilde t_{1,2}} = \{360, 460\}\gev \\
   m_{\tilde Q, \tilde u_R, \tilde d_R} = \{ 495, 460, 455\}\gev && 
  m_{\tilde L, \tilde e_R} = \{ 230, 130 \}\gev  \\
   m_{\chi^0_i} = \{ 60, 110, 410, 420, 1010 \}\gev  &&
   m_{\chi^\pm_i} = \{150, 595\}\gev \\
    m_{A_1^0, H_1^0, H^\pm} = \{510, 515, 515\}\gev && m_{A_2^0, H_2^0} = \{1010,1010\}\gev 
    \end{array} \nonumber
\eeq
Notice that there are now five neutralinos, with the inclusion of the ``singlino", the fermionic component of $S$. A number of sparticles in this particular spectrum are easily discovered at the LHC. However the Higgs bosons, apart from the lightest Higgs, will be a challenge for the LHC and may require a next generation lepton collider. The singlet, which is quite heavy, is well beyond any hope for discovery at the LHC, and so its presence must be felt indirectly, through the mass correction it induces in the lightest Higgs. 

If this spectrum of squarks, sleptons, neutralinos and charginos were to be discovered at the LHC and interpreted in terms of the MSSM, one would predict a light Higgs mass of only $74\gev$, far below the LEP bound. Even if one were to allow for maximal mixing (\ie, large $A_t$, in contradiction to gauge mediation), one could push the Higgs mass up, but only to $102\gev$. Thus, observation of this model at the LHC, with a light Higgs mass above $102\gev$, would require physics beyond the MSSM, and the S-MSSM would have to be a strong candidate for that new physics. Unfortunately, the smoking gun of the S-MSSM, the singlet, could hide itself from direct discovery for many years. 

Of course, the fact that this is a gauge-mediated model does imply some clear experimental effects, including correlations among the sparticle masses and possibly the decay of the NLSP inside the detector. From a cosmological perspective, the gravitino is the stable dark matter candidate, but our singlet sector plays no role in the calculation of the relic abundances.

Finally, although the results shown above assume one pair of messengers ($n=1$ in Eqs.~(\ref{SBgauginos}) -- (\ref{SBsfermions})), we repeated our analysis for $n=2,3$ and found no significant differences between the three cases. Therefore we only show the $n=1$ case here.


\section{\Large Conclusions}

The S-MSSM, which we introduced in Ref.~\cite{smssm}, is simply a generalized version of the NMSSM -- it is the MSSM with a singlet, but also with an explicit $\mu$-term and mass term for the singlet. While most singlet extensions of the MSSM attempt to solve the $\mu$-problem, we are using the singlet to solve the little hierarchy problem. In our previous paper we found Higgs masses as large as $140\gev$ with stops below $1\tev$. As a low-energy effective theory, however, we were free to choose our parameters to fit the problem at hand, namely pushing up the Higgs mass. While we did find models that are free of stop-induced fine tuning using sets of inputs that did not themselves appear to be tuned, it is possible that we were making choices for the parameters that could not be realized in a realistic model of SUSY-breaking, and would therefore be less natural than they appear. 

In this paper, we embedded the S-MSSSM into a gauge-mediated SUSY breaking scheme. Within this scheme, there are very few free parameters, and so the low-energy parameters are highly constrained. Furthermore, because the $A$-terms are generated at higher loops and are thus small, it is difficult to push up the light Higgs mass using the $A_t$-dependent corrections. Thus gauge mediation provides a much more stringent test of the S-MSSM idea. From our results here, it appears that the S-MSSM remains viable once embedded in GMSB, producing Higgs masses up to about $122\gev$ without fine tuning in either the stop sector or anywhere else in the model. The choices of parameters that gave Higgs masses above the LEP bound are quite generic. To a good approximation, in order to obtain Higgs masses above the LEP bound one only requires that $\tan\beta$ is low (between 2 and 6), $\mu_s$ is several times $\mu$ (with $\mu>0$), and that the messenger scale is not too low. Most model points that fit this description are consistent with all direct experimental constraints. And all this is done without resorting to non-perturbative values for $\lambda$, lowering the cut off, or having stop masses in the multi-TeV range, as with some competing models. 

Along the way we found that it is not very difficult to embed the S-MSSM into a gauge-mediated scheme, which is quite unlike the NMSSM. We do not require sizable $m_s^2$ as in the NMSSM, and so non-minimal GMSB embeddings are not necessary. While we do not attempt to explain the source of the $\mu$, $B_\mu$ or $\mu_s$-terms, this is no worse than most gauge-mediated embeddings of the MSSM. This implies, that like the gauge-mediated MSSM, the gauge-mediated S-MSSM is incomplete. Our primary assumption, however, is that any eventual understanding of how $\mu$ and $\mu_s$ are generated within this model will not disrupt the scalar potential at the weak scale.

One of the additional benefits of the S-MSSM is that it opens up parameter space that was previously unavailable in the NMSSM. In particular, the tunings that are required within the NMSSM in order to break the electroweak symmetry while producing a $\mu$-term large enough to pass chargino mass constraints are absent here. The model presented here works for almost any possible value of $A_\lambda$. And although we set $\kappa=A_\kappa=0$ for simplicity, they are both free to take on a wide range of values, with only the constraint that $\kappa\ll 1$ so as not to suppress the perturbative bound on $\lambda$.

In summary, by adding a singlet to the MSSM, but not requiring that it solve the $\mu$-problem, we find a phenomenologically viable, very minimal and quite interesting extension of the MSSM. In particular, this model, the S-MSSM, seems to naturally solve the little hierarchy problem of the MSSM, and this solution is stable to embedding into a gauge-mediated scheme for SUSY breaking. Though the model is almost identical to the MSSM from the perspective of the LHC, the enhancement to the Higgs mass, even when stops are light, would be the first evidence that this model is the correct description of nature at the weak scale.

\section*{Acknowledgments}
AD and CK wish to thank the Aspen Center for Physics where this project was begun. This work was partly supported by the National Science Foundation under grant PHY-0905383-ARRA.

\section*{Appendix}
\renewcommand{\theequation}{A.\arabic{equation}}
\setcounter{equation}{0}  


Here we will give some of the details of the scalar potential and its minimization which were left out of the main body of the paper. The model itself it defined by its superpotential, given in Eq.~(\ref{W}):
\begin{equation}
W_{\rm SMSSM}=W_{\rm Yukawa} + (\mu+\lambda {S}) H_{u}{H_{d}}+\frac{1}{2}\mu_{s}{S}^{2}.
\end{equation}
To this we add the corresponding soft-breaking terms:
\begin{equation}
\label{eq:soft}
V_\text{soft} = m_{H_u}^2 |H_u|^2 + m_{H_d}^2 |H_d|^2 + m_s^2 |S|^2
+ ( B_\mu H_u H_d + h.c. ) + ( A_{\lambda} \lambda S H_u H_d + h.c. )
\end{equation}
The potential for the $H_u$, $H_d$ and $S$ scalar fields is given by the familiar $F$- and $D$-terms, as well as soft breaking operators. Written in full, the scalar potential is given by
\begin{eqnarray}
\label{eq:AppPotential}
V &=& V_F + V_D + V_\text{soft} \nonumber \\
&=& ( m_{H_u}^2 + |\mu + \lambda S|^2 ) | H_u |^2 +
( m_{H_d}^2 + |\mu + \lambda S|^2 ) | H_d |^2 + m_s^2 |S|^2 \nonumber \\
&\indent &
+ | \mu_s S - \lambda H_u H_d |^2
+ ( ( B_\mu + \lambda A_{\lambda} S ) ( H_u H_d ) + B_s S^2 + h.c. ) \nonumber \\ 
&\indent &
+ \frac{1}{8} ( g^2 + g'^2 ) ( |H_u|^2 - |H_d|^2 )^2
+ \frac{1}{2} g^2 |H_u^{\dagger} H_d|^2.
\end{eqnarray} 
Notice that when $S$ gets a vev, it will shift $\mu$ by an amount $\lambda v_s$. By minimizing this potential with respect to the neutral components of the three fields, we obtain:
\begin{eqnarray}
\label{eq:MinConds}
m_{H_u}^{2}+\mu^{2}+\lambda^{2}v_{s}^{2}+\lambda^{2}v^{2}\cos^{2}\beta+2\lambda\mu v_{s}+\frac{1}{4}(g^{2}+g^{\prime 2})(v_{u}^{2}-v_{d}^{2})\quad\quad\quad\quad \nonumber \\
=(B_\mu+(\lambda\mu_{s}+\lambda A_{\lambda})v_{s})\cot\beta, \nonumber \\
m_{H_d}^{2}+\mu^{2}+\lambda^{2}v_{s}^{2}+\lambda^{2}v^{2}\sin^{2}\beta+2\lambda\mu v_{s}-\frac{1}{4}(g^{2}+g^{\prime 2})(v_{u}^{2}-v_{d}^{2}) \quad\quad\quad\quad \label{eq:mins} \\
=(B_\mu+(\lambda\mu_{s}+\lambda A_{\lambda})v_{s})\tan\beta, \nonumber \\
\frac{v^{2}\sin2\beta}{2}(\lambda\mu_{s}+\lambda A_{\lambda})-\lambda\mu v^{2}=(\mu_{s}^{2}+\lambda^{2}v^{2}+m_{s}^{2}+2B_{s})v_{s} \nonumber.~~\,
\end{eqnarray}
where $\left\langle H^{0}_{u}\right\rangle=v_{u} = v \sin\beta, \left\langle H^{0}_{d}\right\rangle=v_{d} = v \cos\beta, \left\langle S \right\rangle = v_s$ and $v^{2}_{u}+v^{2}_{d}=v^{2}$. When these are added and subtracted, one obtains Eqs.~(\ref{eq:m_Z}) -- (\ref{eq:Svev}) in the text.

The mass matrix for the CP-odd components of the neutral Higgs and $S$ scalar fields is symmetric and given by the following terms:
\begin{eqnarray}
M^{2}_{11}&=&(B_\mu+(\lambda\mu_{s}+\lambda A_{\lambda})v_{s})\cot\beta \nonumber \\
M^{2}_{12}&=&B_\mu+(\lambda\mu_{s}+\lambda A_{\lambda})v_{s} \nonumber \\
M^{2}_{13}&=&(-\mu_{s}+A_{\lambda}\lambda)v\cos\beta  \\
M^{2}_{22}&=&(B_\mu+(\lambda\mu_{s}+\lambda A_{\lambda})v_{s})\tan\beta \nonumber \\
M^{2}_{23}&=&(-\mu_{s}+A_{\lambda}\lambda)v\sin\beta \nonumber \\
M^{2}_{33}&=&\mu_{s}^{2}+\lambda^{2}v^{2}+m_{s}^{2}-2B_{s}. \nonumber
\end{eqnarray}
Because this matrix has a zero eigenvalue corresponding to the longitudinal polarization of the $Z^0$, simple closed forms for the eigenvalues are easy to find. However it is more helpful to express the eigenvalues in terms of an expansion in inverse powers of $\mu_s$, in which we assume $\mu_s \gg \mu,v,A_\lambda$:
\begin{eqnarray}
m_{A^{0}_{1}}^{2}&=&\frac{2B_\mu}{\sin2\beta}+\frac{2\lambda^2v^2}{\mu_s}\left(2A_{\lambda}-\frac{\mu}{\sin2\beta}\right)+\cdots \\
m_{A^{0}_{2}}&=&\mu_{s}^{2}+(2\lambda^{2}v^{2}+m_{s}^{2}-2B_{s})+\cdots 
\end{eqnarray}
where the ellipses represent higher orders in $1/\mu_s$. (All calculations given in this paper were done numerically and without use of these expansions.)

The mass matrix for the charged Higgs is only $2\times2$, since the singlet has no charged component, but it still contains singlet-induced corrections:\begin{eqnarray}
M^{2}_{11}&=&(B_\mu+(\lambda\mu_{s}+\lambda A_{\lambda})v_{s})\cot\beta + (m_W^2 - \lambda^2 v^2) \cos^{2}\beta \nonumber \\
M^{2}_{12}&=&B_\mu+(\lambda\mu_{s}+\lambda A_{\lambda})v_{s}+\frac{1}{2} (m_W^2 - \lambda^2 v^2) \sin2\beta \\
M^{2}_{22}&=&(B_\mu+(\lambda\mu_{s}+\lambda A_{\lambda})v_{s})\tan\beta + (m_W^2 - \lambda^2 v^2) \sin^{2}\beta \nonumber 
\end{eqnarray}
using $m_{W}^{2}=g^{2}v^{2}/2$. Besides the massless Goldstone mode, the charged eigenstate has mass:
\begin{eqnarray}
m_{H^{\pm}}^{2}&=&m^2_{A^0_1}+m_{W}^{2}+\frac{2\lambda^2 v^2}{\mu_{s}}\left\{3A_{\lambda}+2\mu\left(\frac{1}{\sin2\beta}-\frac{1}{\sin^2 2\beta}\right)\right\} +\cdots
\end{eqnarray}
Notice that we obtain the usual MSSM result that $m^2_{H^\pm}=m_{A^0}^2+m_W^2$ in the limit $\mu_s\to\infty$, and taking $A^0_1$ to be the usual $A^0$, and so the heavy Higgs doublet decouples in the large $m_{A^0}$ limit, just as in the MSSM.

Finally, the mass matrix for the CP-even Higgs bosons is given by the following terms:
\begin{eqnarray}\label{eq:fr}
M^{2}_{11}&=&(B_\mu+(\lambda\mu_{s}+\lambda A_{\lambda})v_{s})\cot\beta+m^{2}_{Z}\sin^{2}\beta \nonumber \\
M^{2}_{12}&=&-(B_\mu+(\lambda\mu_{s}+\lambda A_{\lambda})v_{s})-m^{2}_{Z}\sin\beta\cos\beta+2\lambda^{2}v^{2}\sin\beta\cos\beta \nonumber \\
M^{2}_{13}&=&2\lambda^{2}v_{s}v\sin\beta-(\lambda\mu_{s}+\lambda A_{\lambda})v\cos\beta+2\lambda\mu v\sin\beta \\
M^{2}_{22}&=&(B_\mu+(\lambda\mu_{s}+\lambda A_{\lambda})v_{s})\tan\beta+m^{2}_{Z}\cos^{2}\beta \nonumber \\
M^{2}_{23}&=&2\lambda^{2}v_{s}v\cos\beta-(\lambda\mu_{s}+\lambda A_{\lambda})v\sin\beta+2\lambda\mu v\cos\beta \nonumber \\
M^{2}_{33}&=&\mu_{s}^{2}+\lambda^{2}v^{2}+m_{s}^{2}+2B_{s} \nonumber
\end{eqnarray}
The full expressions for the eigenvalues of this $3 \times 3$ matrix are much too long to include here, but if we assume again that the singlet mass is large and expand in inverse powers of $\mu_s$, then the eigenvalues take on a simple form:
\begin{eqnarray}
m^{2}_{h^{0}}&=&m^{2}_{h^{0}_{\rm MSSM}}+\frac{\lambda^{2}v^{2}}{\mu_{s}}(\mu\sin2\beta-A_{\lambda}-\Delta)+\cdots  \\
m^{2}_{H^{0}_{1}}&=&m^{2}_{H^{0}_{\rm MSSM}}+\frac{\lambda^{2}v^{2}}{\mu_{s}}(\mu\sin2\beta-A_{\lambda}+\Delta)+\cdots  \\
m^{2}_{H^{0}_{2}}&=&\mu_{s}^{2}+(2\lambda^{2}v^{2}+m_{s}^{2}-2B_{s})+\cdots  
\end{eqnarray}
where the terms labelled ``MSSM" are the scalar Higgs boson masses obtained from Eq.~\ref{eq:AppPotential} with $\lambda=\mu_{s}=B_s=A_\lambda=0$. The mass term $\Delta$ corresponds to the $1/{\mu_{s}}$ correction to the splitting of the scalar Higgs boson masses and it is given by
\begin{equation}
\Delta \equiv 
\frac{A_{\lambda}(m^{2}_{Z}-m^{2}_{A_1^0})\cos^{2}2\beta-\mu(m^{2}_{A_1^0}+m^{2}_{Z})\sin2\beta}{\sqrt{(m^{2}_{A_1^0}+m^{2}_{Z})^{2}-4m^{2}_{A_1^0}m^{2}_{Z}\cos^{2}2\beta}}.
\end{equation}
In the text we also presented a form for the CP-even Higgs masses, Eq.~(\ref{eq:largemA}), in which we have included the next order in $1/\mu_s$ while taking $m_{A_1^0}$ large.

~

Among the parameters that enter the scalar potential, a special role is played by $\lambda$, the coupling of the singlet to the usual Higgs doublets. In this analysis, we always take $\lambda$ to be maximal while preserving the unification of the gauge couplings at $\sim 10^{16}\gev$, a constraint often called perturbative unification. The maximal value of $\lambda$ must be obtained by running the renormalization group equations (RGEs) for $\lambda$, the Yukawa couplings and the gauge couplings between the weak and high scales. We did this at two loops for each coupling.

The RGEs used in the analysis are the usual RGEs of the NMSSM, though with $\kappa=0$~\cite{2loops}. Rather than reproduce all the relevant RGEs here, we will refer the reader to that reference. However, the RGE for $\lambda$ itself is worth presenting:
\begin{eqnarray} 
\frac{d \lambda}{dt} &=& 
	\frac{\lambda}{16 \pi^2 } \Big( 3 (y_t^2 + y_b^2) + y_\tau^2 
		+ 4 \lambda^2 - \frac{3}{5} g_1^2 - 3 g_2^2 \Big)   \\ \nonumber
	&+&
	\frac{\lambda}{ (16 \pi^2)^2 } 
		\Big( \frac{207}{50} g_1^4 + \frac{15}{2} g_2^4 + 
   \frac{9}{5} g_1^2 g_2^2 + \frac{6}{5} g_1^2 \lambda ^2 + 
   \frac{6}{5} g_1^2 y_\tau^2   \\ \nonumber
	&& \hskip 1cm + \frac{4}{5} g_1^2 y_t^2 - 
   \frac{2}{5} g_1^2 y_b^2 + 6 g_2^2 \lambda ^2 + 
   16 g_3^2 (y_t^2 + y_b^2)  -6 y_t^2 y_b^2 \\ \nonumber
	&& \hskip 1cm  - 9 (y_t^4 + y_b^4) - 3 y_\tau^4 - 
   10 \lambda ^4 - \lambda^2 (9 y_t^2 + 9 y_b^2 + 3 y_\tau^2) \Big) \nonumber
\eea
where $g_1=\sqrt{5/3} g'$ is the usual GUT-normalized hypercharge coupling, $g_2 = g$ and $t \equiv \log Q$ for arbitrary scale $Q$. 

The leading terms in the RGE for $\lambda$ are the contributions of top Yukawa coupling and $\lambda$ itself:
\beq
\frac{d\lambda}{dt} \simeq \frac{\lambda}{16\pi^2} (3y_t^2+4\lambda^2).
\eeq
Both terms tend to increase $\lambda$ in the ultraviolet and therefore the maximum value of $\lambda$ depends sensitively on the value of $y_t=m_t/(v\sin\beta)$. Thus the maximum value of $\lambda$ in the infrared, $\lambda_{\rm max}$, depends sensitively on $\tan\beta$. This behavior is evident in Figure~\ref{fig:MaxLambda}, where we have plotted the maximum value of $\lambda$ as a function of $\tan\beta$. Here we define numerically the maximum value of $\lambda(\mw)$ as that value which runs to $\lambda=\pi$ at the gauge unification scale. (The exact value we choose for $\lambda$ at the unification scale has almost no effect on our results.)

\begin{figure}[t]
\begin{center}
\includegraphics[width=0.7\linewidth]{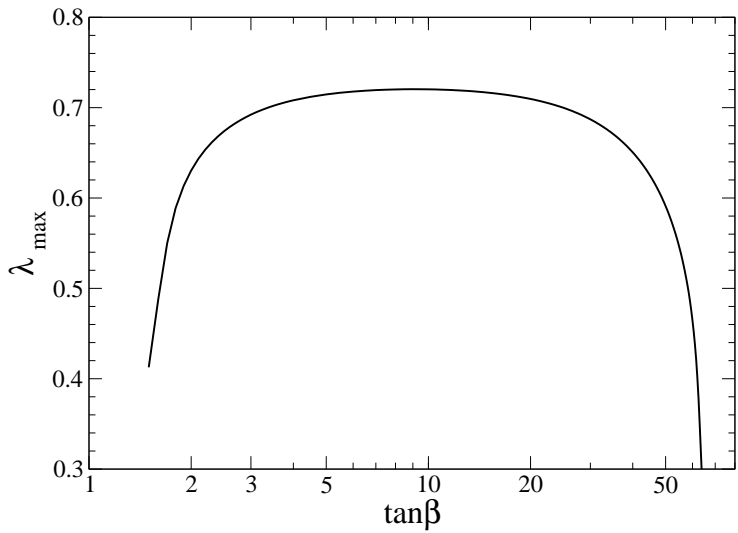}
\caption{The maximum value of $\lambda$ at the weak scale, consistent with perturbativity of all couplings up to the gauge unification scale, as a function of $\tan\beta$. \label{fig:MaxLambda}}
\end{center}
\end{figure}

Looking at Figure~\ref{fig:MaxLambda}, we see that for $\tan\beta\simeq 1$, perturbativity breaks down before the unification scale for any value of $\lambda$, but for $\tan\beta\gsim 1.5$, non-zero values of $\lambda$ are allowed. In fact, the maximum value of $\lambda$ rises quickly at low $\tan\beta$ until it plateaus at $\lambda\simeq 0.7$ for $2.5\lsim\tan\beta\lsim 40$. At very high $\tan\beta$, the bottom Yukawa suddenly becomes important and $\lambda_{\rm max}$ is again suppressed. Throughout this analysis we take the value of $\lambda_{\rm max}$ appropriate for the value of $\tan\beta$ being studied.

One last note about taking $\lambda=\lambda_{\rm max}$ in our analysis. While this may appear to be a tuning, it is in fact quite natural. The value of $\lambda_{\rm max}$ is actually an infrared pseudo-fixed point of the RGEs -- if we choose $\lambda$ to be any large, but perturbative, value at the high scale, it will run down to a value very close to $\lambda_{\rm max}$ in the infrared, regardless of the exact value at the high scale. This is a similar behavior to what is observed for the top Yukawa, and in this sense it is equally natural as a large $y_t$.

\bibliographystyle{unsrt}

\end{document}